\documentstyle[12pt]{article}
\def \b{{\cal B}}
\def\bea{\begin{eqnarray}}
\def\eea{\end{eqnarray}}
\def \beq{\begin{equation}}
\def \eeq{\end{equation}}
\def \ite{{\it et al.}}
\def \s{\sqrt{2}}

\begin{document}
\renewcommand{\thetable}{\Roman{table}}
\rightline{TECHNION-PH-99-39}
\rightline{November 1999}
\rightline{hep-ph/9911429}
\bigskip

\centerline{\bf ELECTROWEAK PENGUIN AMPLITUDES AND CONSTRAINTS} 
\centerline{\bf ON $\gamma$ IN CHARMLESS $B\to VP$ DECAYS
\footnote{to be published in Phys.~Rev.~D.}}
\bigskip\bigskip
\centerline{\it Michael Gronau}
\centerline{\it Physics Department}
\centerline{\it Technion -- Israel Institute of Technology, 32000 Haifa,
Israel}

\bigskip

\centerline{\bf ABSTRACT}
\bigskip

\begin{quote}
Electroweak penguin (EWP) amplitudes are studied model-independently 
in $B$ meson decays to charmless final states consisting of a vector meson 
($V$) and a pseudoscalar meson ($P$). A set of SU(3) relations is derived 
between EWP contributions and tree amplitudes, in the approximation of 
retaining only the dominant EWP operators $Q_9$ and $Q_{10}$ . Two applications 
are described for constraining the weak phase $\gamma$, in $B^{\pm}\to \rho^{\pm} K^0$ and 
$B^{\pm}\to \rho^0 K^{\pm}$ (or $B^{\pm}\to K^{*\pm}\pi^0$ and $B^{\pm}\to K^{*0}
\pi^{\pm}$), and in $B^0\to K^{*\pm}\pi^{\mp}$ and $B^{\pm}\to\phi K^{\pm}$.
Theoretical uncertainties are discussed.
\end{quote}

\bigskip
\leftline{PACS numbers: 13.25.Hw, 14.40.Nd, 11.30.Er, 12.15.Ji}
\bigskip

\centerline{\bf I.  INTRODUCTION}
\bigskip

$B$ meson decays to charmless final states open a window into new phenomena
of CP violation \cite{MGrev}, providing useful information about the 
Kobayashi-Maskawa phase $\gamma={\rm Arg} V^*_{ub}$. Decays to states 
consisting of two light pseudoscalars ($B\to PP$), such as $B\to \pi\pi$ and 
$B\to K\pi$, have been for some time the subject of extensive studies. The 
amplitudes of these processes 
involve hadronic matrix elements of a low energy effective weak 
Hamiltonian between an initial $B$ meson and two final pseudoscalar mesons. 
The weak Hamiltonian consists of the sum of three types of four quark 
operators,
namely two $(V-A)(V-A)$ current-current operators ($Q_{1,2}$), four QCD 
penguin 
operators ($Q_{3,4,5,6}$), and four electroweak penguin (EWP) operators 
($Q_{7,8,9,10}$) with different chiral structures. 
A major line of analysis \cite{SU(3)} consists of model-independent studies of 
hadronic matrix elements of these operators, which do not rely on 
factorization-based models \cite{FACT,BBNS}. Approximate flavor SU(3) symmetry 
of strong interactions 
was employed \cite{Zepp,Chau,GHLR,Grin} in order to describe these matrix 
elements in a 
graphical manner in terms of a relatively small number of amplitudes. 

A useful simplification was achieved \cite{NRPL,GPY} by noting that in certain 
cases, such as in $B$ decay to an isospin $3/2~K\pi$ state, the {\it dominant} 
EWP amplitude is simply related by SU(3) to the corresponding 
current-current contribution, and does not introduce a new unknown quantity 
into 
the analysis. This simplification, obtained when retaining only the 
$(V-A)(V-A)$ 
EWP operators ($Q_9$ and $Q_{10}$), led to a promising way of 
measuring the weak phase $\gamma$ \cite{NRPRL,FLE}. 

A first SU(3) analysis of $B$ mesons decays to a charmless vector meson ($V$) 
and a pseudoscalar meson ($P$), classifiying contributions in terms of 
graphical 
SU(3) amplitudes, was presented in \cite{VP1}. Several 
factorization-based calculations of these processes can be found in 
\cite{FACT}.
Measurements of $B\to VP$ decays were reported by the CLEO collaboration 
working at the Cornell Electron Storage Ring (CESR) \cite{CLEO}.
The experimental results were used recently \cite{VP2} in order to identify 
dominant and subdominant interfering amplitudes in certain processes.  
Interference effects between these amplitudes seem to favor 
(but do not necessarily imply) a weak phase 
$\gamma$ in the second quadrant of the unitarity triangle plot. A similar 
conclusion was drawn recently in factorization-based analyses \cite{HOU}.
Such values of $\gamma$ are in sharp conflict with an overall CKM parameter 
analysis \cite{PRS} based on very optimistic assumptions about theoretical 
uncertainties in hadronic parameters. A more conservative estimate of 
these errors \cite{PS} implies $\gamma\le 90^{\circ}$, based primarily on 
a lower limit for $B_s-\bar B_s$ mixing, $\Delta m_s > 14.3~ps^{-1}$ 
\cite{Blai} . Using a somewhat wider range for the relevant SU(3) breaking 
parameter, $f_{B_s}\sqrt{B_{B_s}}/f_B\sqrt{B_B}$, we concluded recently 
\cite{VP2} that values of $\gamma$ slighly larger than $90^{\circ}$ cannot be 
definitely excluded.   

Two major assumptions were made in \cite{VP2} in order to arrive at the 
indication that 
$\gamma>90^\circ$ within the framework of flavor SU(3).
1. The relative strong phase between penguin and tree amplitudes in 
$B^0\to K^{*+}\pi^-$ was assumed to be smaller than $90^\circ$. 2. The 
magnitude 
of a color-favored EWP contribution in $B^+ \to \phi K^+$ was 
taken from factorization-based calculations \cite{RFDH,ALI}, and 
color-suppressed 
EWP contributions were neglected. The first assumption is rather 
plausible, and can be justified on the basis of both perturbative 
{\cite{BBNS,BSS} and statistical \cite{SW} estimates of final state phases. 
The second assumption is 
manifestly model-dependent. One would hope to be able to replace it by 
a model-independent study.

In order to study $B\to VP$ decays in a model-independent manner, we propose 
in this article to derive SU(3) relations between
color-favored and color-suppressed EWP amplitudes, on the one 
hand, and current-current contributions on the other hand. The 
relations, obtained in the approximation of retaining only the dominant 
$(V-A)(V-A)$ EWP operators, eliminate eight of the 
hadronic parameters describing charmless $B\to VP$ decays in the SU(3) 
framework.  

In Section II we recall the general SU(3) structure of the effective weak 
Hamiltonian, paying particular attention to its current-current part and 
its dominant electroweak term. We show that corresponding components of these 
operators, transforming as given SU(3) representations, are proportinal
to each other. In Sec. III we use this feature to study the SU(3) structure of 
$B\to VP$ amplitudes, and in Sec. IV we express EWP contributions in $B\to VP$ 
decays in terms of corresponding tree amplitudes. 

Two applications are 
demonstrated in Sec. V for constraining the weak phase $\gamma$ by 
charge-averaged ratios of rates in $B^{\pm}
\to \rho^{\pm} K^0$ and $B^{\pm}\to \rho^0 K^{\pm}$ (or $B^{\pm}\to K^{*\pm}
\pi^0$ and $B^{\pm}\to K^{*0} \pi^{\pm}$), and in $B^0\to K^{*\pm}\pi^{\mp}$
and $B^{\pm}\to\phi K^{\pm}$. In the first case, no data exist at this time.
In the second case, existing data may imply a lower limit on 
$\gamma$, provided that a better understanding is achieved for SU(3) breaking
in QCD penguin amplitudes and for the effects of color- and OZI-suppressed 
amplitudes. We conclude in Sec. VI. 

\bigskip

\centerline{\bf II. SU(3) STRUCTURE OF WEAK HAMILTONIAN}
\bigskip

The low energy effective weak Hamiltonian governing $B$ meson decays is given 
by \cite{BBL}
\beq\label{H}
{\cal H} = \frac{G_F}{\sqrt2}
\sum_{q=d,s}\left(\sum_{q'=u,c} \lambda_{q'}^{(q)}
[c_1 (\bar bq')_{V-A}(\bar q'q)_{V-A} + c_2 (\bar bq)_{V-A}(\bar q'q')_{V-A}]
-  \lambda_t^{(q)}\sum_{i=3}^{10}c_i Q^{(q)}_i\right)~,
\eeq
where $\lambda_{q'}^{(q)}=V_{q'b}^*V_{q'q},~q=d,s,~q'=u,c,t,~\lambda_u^{(q)}+
\lambda_c^{(q)}+\lambda_t^{(q)}=0$. The first terms, involving the coefficients 
$c_1$ and $c_2$ and describing both $\bar b\to\bar q u\bar u$ and $\bar b\to
\bar q c\bar c$, will be referred to as ``current-current" operators, 
while the other terms, involving $c_i~i=3-10$, consist of four QCD 
penguin operators ($i=3-6$) and four EWP operators ($i=7-10$). 
The EWP operators with the dominant Wilson
coefficients, $Q_9$ and $Q_{10}$, both have a $(V-A)(V-A)$ structure similar to 
the current-current term. Their flavor structure is 
\bea\label{EWPen}
Q^{(q)}_9 &=& \frac32\left[
(\bar bq)(\frac23\bar uu-\frac13\bar dd -\frac13\bar ss +\frac23\bar cc)
\right]~,\nonumber\\
Q^{(q)}_{10} &=& \frac32\left[
\frac23(\bar bu)(\bar uq)-\frac13(\bar bd)(\bar dq)-\frac13(\bar bs)(\bar 
sq)+\frac23(\bar bc)(\bar cq)\right]~.
\eea

All four-quark operators appearing in (\ref{H})
are of the form $(\bar bq_1)(\bar q_2 q_3)$ and can be written as a sum of 
${\overline {\bf 15}}$, ${\bf 6}$ and ${\overline {\bf 3}}$, into which the 
product ${\overline{\bf 3}}\otimes {\bf 3}\otimes {\overline{\bf 3}}$ is 
decomposed 
\cite{Zepp,GHLR,Grin}. The representation ${\overline{\bf 3}}$ appears both 
symmetric (${\overline{\bf 3}}^{(s)}$), and antisymmetric 
(${\overline{\bf 3}}^{(a)}$) 
under the interchange of $q_1$ and $q_3$. Four-quark operators, belonging to 
each of these SU(3) representations and carrying given values of isospin, are 
listed in the appendix of \cite{GPY}. 

The ``tree" part of the current-current Hamiltonian, corresponding to the term
$q'=u$ describing $\bar b\to\bar q u\bar u$ transitions, can be written as 
\cite{GPY}
\bea\label{T}
\frac{\s{\cal H}_T}{G_F} = -\lambda_u^{(s)}
[\frac{c_1-c_2}{2}(\overline{\bf 3}^{(a)}_{I=0} + {\bf 6}_{I=1}) +
\frac{c_1+c_2}{2}(\overline {\bf 15}_{I=1} + \frac{1}{\sqrt2}\overline 
{\bf 15}_{I=0} - \frac{1}{\sqrt2}\overline{\bf 3}^{(s)}_{I=0}) ]\nonumber\\
- \lambda_u^{(d)}
[\frac{c_1-c_2}{2}(\overline{\bf 3}^{(a)}_{I=\frac12} - {\bf 6}_{I=\frac12}) +
\frac{c_1+c_2}{2}(\frac{2}{\sqrt3}\overline {\bf 15}_{I=\frac32} + 
\frac{1}{\sqrt 6}\overline {\bf 15}_{I=\frac12}
-\frac{1}{\sqrt2}\overline{\bf 3}^{(s)}_{I=\frac12})]~.
\eea
The dominant EWP term, excluding $\bar b\to\bar q c\bar c$, 
(to be referred to as the noncharming EWP operator), is 
\bea\label{EWP}
&&\frac{\s{\cal H}_{EWP}}{G_F} =\\
&&-\frac{3\lambda_t^{(s)}}{2}[
\frac{c_9-c_{10}}{2}(\frac{1}{3}\overline{\bf 3}^{(a)}_{I=0} + {\bf 6}_{I=1}) +
\frac{c_9+c_{10}}{2}(-\overline {\bf 15}_{I=1} 
-\frac{1}{\sqrt2}\overline {\bf 15}_{I=0}
-\frac{1}{3\sqrt2}\overline{\bf 3}^{(s)}_{I=0} )]\nonumber\\
&&-\frac{3\lambda_t^{(d)}}{2}[\frac{c_9-c_{10}}{2}(\frac{1}{3}
\overline{\bf 3}^{(a)}_{I=\frac12} - {\bf 6}_{I=\frac12}) +
\frac{c_9+c_{10}}{2}(-\frac{2}{\sqrt 3}\overline {\bf 15}_{I=\frac32} -
\frac{1}{\sqrt 6}\overline {\bf 15}_{I=\frac12}
-\frac{1}{3\sqrt2}\overline{\bf 3}^{(s)}_{I=\frac12})].\nonumber
\eea

Eqs.~(\ref{T}) and (\ref{EWP}) teach us something very important. 
{\it For a given strangeness-change, the $\overline {\bf 15}$ and ${\bf 6}$ 
components of the tree operator and the dominant EWP operator 
in the Hamiltonian are proportional to each other}:
\bea\label{15}
{\cal H}^{(q)}_{EWP}(\overline{\bf 15}) &=& -\frac32 \frac{c_9+c_{10}}{c_1+c_2} 
\frac{\lambda_t^{(q)}}{\lambda_u^{(q)}} 
{\cal H}^{(q)}_T(\overline{\bf 15})~,\\
\label{6}
{\cal H}^{(q)}_{EWP}({\bf 6}) &=& \frac32 \frac{c_9-c_{10}}{c_1-c_2}
\frac{\lambda_t^{(q)}}{\lambda_u^{(q)}}{\cal H}^{(q)}_T({\bf 6})~.
\eea
Here the superscripts $q=d,s$ denote strangeness-conserving and 
strangeness-changing transitions, respectively. The above two relations are 
unaffected by the inclusion of the current-current and EWP operators
describing $\bar b\to\bar q c\bar c$ transitions, each of which transforms as 
an antitriplet.

A similar relation between the $\overline{\bf 3}$ parts of ${\cal H}_{EWP}$ and 
${\cal H}_T$ holds only when the two ratios of Wilson coefficients, 
$(c_9+c_{10})/(c_1+c_2)$ and $(c_9-c_{10})/(c_1-c_2)$,  
are equal. Indeed, these two ratios, which are approximately renormalization 
scale independent, are equal to a very good approximation \cite{GPY}. At a 
scale $\mu=m_b$, they differ by less than 3$\%$ \cite{BBL,MN} 
\beq
\frac{c_9+c_{10}}{c_1+c_2} = -1.139\alpha~,\qquad
\frac{c_9-c_{10}}{c_1-c_2} = -1.107\alpha~,
\eeq
where $\alpha=1/129$. We will take the average of the two ratios and denote it 
by $\kappa$
\beq
\kappa \equiv \frac{c_9+c_{10}}{c_1+c_2} = \frac{c_9-c_{10}}{c_1-c_2} = 
-1.12\alpha~.
\eeq
In this approximation, we also have 
\beq\label{3}
{\cal H}^{(q)}_{EWP}(\overline{\bf 3}) = \frac12 \kappa \frac{\lambda_t^{(q)}}
{\lambda_u^{(q)}} 
{\cal H}^{(q)}_T(\overline{\bf 3})~.
\eeq
We note that this relation excludes the current-current and EWP $\bar b\to\bar 
q c\bar c$ operators
which transform as an independent antitriplet. As mentioned, the two relations 
(\ref{15}) and ({6}) are unaffected by the inclusion of these operators.

The operator relations (\ref{15}), (\ref{6}) and (\ref{3}) lead to 
corresponding relations between tree and EWP 
amplitudes contributing to various processes. An interesting example 
\cite{NRPL,GPY} is $B\to (K\pi)_{I=3/2}$, in which one chooses the final 
$K\pi$ state to be in $I=3/2$. This S-wave state, which is symmetric under 
interchanging the two SU(3) octets, is pure ${\bf 27}$. 
The only SU(3) operator in the Hamiltonian which contributes to this transition 
is $\overline{\bf 15}$ \cite{Zepp,GHLR,Grin}. Consequently, the ratio of EWP 
and tree contributions in $B\to (K\pi)_{I=3/2}$ is given simply by 
$-(3/2)\kappa(\lambda_t^{(s)}/\lambda_u^{(s)})$. This feature was shown to have 
a useful implication when studying the weak phase $\gamma$ in $B^+\to K\pi$ 
decays. In the next two sections we will study generalizations of this relation 
in $B\to VP$ decays.

\bigskip
\centerline{\bf III. SU(3) DECOMPOSITION OF $B\to VP$ AMPLITUDES}
\bigskip

In Ref.~\cite{VP1,VP2} $B\to VP$ amplitudes were expressed in terms of 
reduced SU(3) amplitudes depicted in graphical form. For the most part, we 
will consider in this paper decay amplitudes into states involving two octet 
mesons, which consist of  $T_M$ (tree), $C_M$ (color-suppressed), $P_M$ 
(QCD-penguin), 
$E_M$ (exchange), $A_M$ (annihilation) and $PA_M$ (penguin annihilation). The 
suffix $M~=~P,~V$ on the three amplitudes $T, C$ and $P$ denotes whether 
the spectator quark is included in a pseudoscalar or vector meson, 
respectively. In $E_M, A_M$ and $PA_M$ the suffix denotes the type of meson
into which the outgoing quark $q_3$ enters in $\bar b q_1\to \bar q_2 q_3$.
In the last six amplitudes the spectator quark enters into the decay 
Hamiltonian. These contributions, which were neglected in \cite{VP1,VP2},  
may be important in the presence of rescattering \cite{resc}, 
and will not be neglected here.  

We will use a somewhat different notation for the graphical amplitudes than in 
\cite{VP1,VP2} by factoring out Cabibbo-Kobayashi-Maskawa (CKM) elements. 
We will also separate the charming penguin contributions \cite{Ciu}, related
to $\bar b\to\bar q c\bar c$, from the noncharming terms. Thus, for instance, 
a typical $\Delta S=1$ amplitude is given by
\bea\label{amp}
A(B^0\to \rho^- K^+) = \lambda^{(s)}_u [-P_{uc,V} - T_V]  + \lambda^{(s)}_t 
[-P_{tc,V} &+& EW(B^0\to \rho^- K^+)\nonumber\\ 
&+& EW_c(B^0\to \rho^- K^+)]~,
\eea
where $P_{uc,V}=P_{u,V}-P_{c,V},~P_{tc,V}=P_{t,V}-P_{c,V}$.
Both $T_V$ and $P_{u,V}$ are contributions from the tree Hamiltonian
(\ref{T}), and will be referred to as tree amplitudes, in spite of the
fact that $P_{u,V}$ may be depicted as a penguin diagram with an internal 
$u$ quark. ~$P_{c,V}$ and $EW_c$ originate 
in $\bar b\to\bar c q\bar c$ current-current and EWP operators, respectively, 
and will be referred to as charming penguin and charming EWP terms. 
Finally, $P_{t,V}$ and 
$EW$ are contributions from QCD penguin operators and from the dominant 
noncharming EWP Hamiltonian (\ref{EWP}), and will be referred to as 
noncharming penguin amplitudes. (One expects $|EW_c|\ll |EW|$.)

In previous analyses, EWP contributions \cite{EWP} multiplying
$\lambda^{(q)}_t$  were taken to be independent of the other terms. 
They were introduced through the substitution \cite{GHLRP} 
\beq
T_M \to t_M \equiv T_M + EW^C_M~~,~~~
C_M \to c_M \equiv C_M + EW_M~,~~~
P_M \to p_M \equiv P_M - {1 \over 3}EW^C_M~.
\eeq
The color-favored ($EW_M$) and color-suppressed ($EW^C_M$)  
EWP amplitudes, in which the spectator enters the meson $M$, were considered 
to be independent of the other amplitudes. They are 
calculable in specific models based on factorization \cite{FACT}.
Four other EWP contributions \cite{EWE}, in which the 
spectator quark enters into the effective EWP Hamiltonian, were neglected. 
Such amplitudes can be enhanced by rescattering. Including these amplitudes 
introduces a total of eight additional unknown parameters into the SU(3) 
analysis.

Here we wish to use the approximate operator equalities (\ref{15})(\ref{6}) 
and (\ref{3}) in order to relate all eight EWP parameters to tree amplitudes.
We will find relations between the dominant noncharming 
EWP contributions $EW$ multiplying $\lambda^{(q)}_t$
and the tree amplitudes multiplying $\lambda^{(q)}_u$. This program, analogous 
to the study of EWP amplitudes in $B\to PP$ decays \cite{GPY}, 
can be carried out by expressesing tree and EWP amplitudes in 
terms of reduced SU(3) matrix elements. 

Counting the number of reduced matrix elements for $B$ decays to two octet $VP$
states, one finds \cite{Zepp} five amplitudes for SU(3) symmetric $VP$ states
\beq\label{sym}
\langle {\bf 1}|\!| {\overline{\bf 3}} |\!| {\bf 3}\rangle~,~~~
\langle {\bf 8_s}|\!| {\overline{\bf 3}} |\!| {\bf 3}\rangle~,~~~
\langle {\bf 8_s}|\!| {\bf 6} |\!| {\bf 3}\rangle~,~~~
\langle {\bf 8_s}|\!| {\overline{\bf 15}} |\!| {\bf 3}\rangle~,~~~
\langle {\bf 27}|\!| {\overline{\bf 15}} |\!| {\bf 3}\rangle~,
\eeq
and five matrix elements for antisymmetric states
\beq\label{antisym}
\langle {\bf 8_a}|\!| {\overline{\bf 3}} |\!| {\bf 3}\rangle~,~~~
\langle {\bf 8_a}|\!| {\bf 6} |\!| {\bf 3}\rangle~,~~~
\langle {\bf 10}|\!| {\bf 6} |\!| {\bf 3}\rangle~,~~~
\langle {\bf 8_a}|\!| {\overline{\bf 15}} |\!| {\bf 3}\rangle~,~~~
\langle {\overline {\bf 10}}|\!| {\overline{\bf 15}} |\!| {\bf 3}\rangle~.
\eeq

The decomposition of $B\to VP$ amplitudes in terms of these reduced matrix
elements, occuring both in tree and EWP contributions, is
given in Table I for $\Delta S=1$ decays. The coefficients of the reduced matrix
elements are tabulated for all four $B\to \rho K$ decay processes. The 
amplitudes 
for $B\to K^*\pi$ processes are obtained by interchanging the SU(3) flavors of 
the vector and pseudoscalar mesons. Consequently, the coefficients of the five 
symmetric elements (\ref{sym}) are the same as in the corresponding 
$B\to \rho K$ decays, whereas the coefficients of the five antisymmetric 
elements (\ref{antisym}) change sign.
 
\begin{table}
\caption{Decomposition of $B\to \rho K$ amplitudes into SU(3) reduced matrix 
elements (12) and (13). The coefficients for corresponding
$B\to K^*\pi$ decays are the same for (12) and have opposite signs for
(13).}
\begin{center}
\begin{tabular}{c c c c c} \hline \hline
Reduced amplitude  & $B^+\to\rho^+K^0$ & $B^+\to\rho^0 K^+$ & 
$B^0\to\rho^-K^+$ & $B^0\to\rho^0K^0$ \\
\hline
$\langle {\bf 1}|\!| {\overline{\bf 3}} |\!| {\bf 3}\rangle$ & 0 & 0 & 0 & 0 
\\
$\langle {\bf 8_s}|\!| {\overline{\bf 3}} |\!| {\bf 3}\rangle$ & $-\sqrt{3/5}$ 
& $\sqrt{3/10}$ & $\sqrt{3/5}$ &  $-\sqrt{3/10}$ \\
$\langle {\bf 8_s}|\!| {\bf 6} |\!| {\bf 3}\rangle$ & $1/\sqrt5$ 
& $-1/\sqrt{10}$ & $1/\sqrt5$ & $-1/\sqrt{10}$ \\
$\langle {\bf 8_s}|\!| {\overline{\bf 15}} |\!| {\bf 3}\rangle$ & 
$-3\sqrt3/5$ & 
$3\sqrt6/10$ & $-\sqrt3/5$ & $\sqrt6/10$ \\
$\langle {\bf 27}|\!| {\overline{\bf 15}} |\!| {\bf 3}\rangle$ & $2\sqrt3/5$ & 
$4\sqrt6/5$ & $4\sqrt3/5$ & $3\sqrt6/5$ \\
$\langle {\bf 8_a}|\!| {\overline{\bf 3}} |\!| {\bf 3}\rangle$ & $1/\sqrt3$ 
& $-1/\sqrt6$ & $-1/\sqrt3$ & $1/\sqrt6$ \\
$\langle {\bf 8_a}|\!| {\bf 6} |\!| {\bf 3}\rangle$ & $-1/3$ & 
$\sqrt2/6$ & $-1/3$ & $\sqrt2/6$ \\
$\langle {\bf 10}|\!| {\bf 6} |\!| {\bf 3}\rangle$ & $\sqrt2/3$ & 
$2/3$ & $\sqrt2/3$ & $2/3$ \\
$\langle {\bf 8_a}|\!| {\overline{\bf 15}} |\!| {\bf 3}\rangle$ 
& $\sqrt{3/5}$ & $-\sqrt{3/10}$ & $1/\sqrt{15}$ & $-1/\sqrt{30}$ \\
$\langle {\overline {\bf 10}}|\!| {\overline{\bf 15}} |\!| {\bf 3}\rangle$ 
& 0 & 0 & $2/\sqrt3$ & $-\sqrt{2/3}$ \\
\hline
\hline
\end{tabular}
\end{center}
\end{table}
 
Expressions of the reduced elements (\ref{sym}) in terms of graphical tree 
amplitudes in $B\to PP$ were given in the appendix of \cite{GHLR}. They can be 
transcribed to the case of $B\to VP$ by defining combinations of amplitudes, 
$X_s\equiv (X_V + X_P)/2$, which are symmetric under interchanging the vector 
and pseudoscalar mesons [we define $(X+Y)_s\equiv X_s+Y_s$]: 
\bea\label{Sreduced}
\langle {\bf 27}|\!| {\overline{\bf 15}} |\!| {\bf 3}\rangle &=& -
\frac{1}{2\sqrt3}(T+C)_s~\\
\langle {\bf 8_s}|\!| {\overline{\bf 15}} |\!| {\bf 3}\rangle &=& -
\frac{1}{8\sqrt3}(T+C+5A+5E)_s~,\\
\langle {\bf 8_s}|\!| {\bf 6} |\!| {\bf 3}\rangle &=& -\frac{\sqrt5}{4}
(T-C-A+E)_s~,\\
\langle {\bf 8_s}|\!| {\overline{\bf 3}} |\!| {\bf 3}\rangle &=& -\frac{1}{8}
\sqrt{\frac{5}{3}}(3T+3A-C-E+8P_u)_s~,\\
\langle {\bf 1}|\!| {\overline{\bf 3}} |\!| {\bf 3}\rangle &=& \frac{1}{2\sqrt3}
(3T-C+8E+8P_u+12PA_u)_s~.
\eea
The set of six graphical amplitudes on the right-hand-sides is over-complete. 
The physical processes involve only five linear combinations of these amplitudes.
Similar relations are obtained for the amplitudes (\ref{antisym}) in terms of
antisymmetric combinations $X_a\equiv (X_V - X_P)/2$ [we define $(X+Y)_a\equiv 
X_a+Y_a$]:
\bea
\langle {\overline {\bf 10}}|\!| {\overline{\bf 15}} |\!| {\bf 3}\rangle &=& \frac{1}
{2\sqrt3}(T+C)_a~,\\
\langle {\bf 8_a}|\!| {\overline{\bf 15}} |\!| {\bf 3}\rangle &=& -\frac{1}{8}
\sqrt{\frac{5}{3}}(T+C-3A-3E)_a~,\\
\langle {\bf 10}|\!| {\bf 6} |\!| {\bf 3}\rangle &=& -\frac{1}{\sqrt2}
(T-C)_a~,\\
\langle {\bf 8_a}|\!| {\bf 6} |\!| {\bf 3}\rangle &=& -\frac{1}{4}
(T-C+3A-3E)_a~,\\ \label{Areduced}
\langle {\bf 8_a}|\!| {\overline{\bf 3}} |\!| {\bf 3}\rangle &=& \frac{\sqrt3}{8}
(3T+3A-C-E+8P_u)_a~.
\eea
Here the number of graphical amplitudes is identical to that of the reduced 
SU(3) matrix elements. The amplitude $PA_{u,a}$ vanishes, since the penguin 
annihilation graph leads to an SU(3) singlet state. 
By substituting the expressions of Eqs.~(\ref{Sreduced})--(\ref{Areduced}) into
Table I, it is straightforward to check that one obtains the appropriate 
graphical description of tree amplitudes for all $B\to PV$ decays, such 
as written directly for $B^0\to\rho^-K^+$ in Eq.~(\ref{amp}). 

\bigskip
\centerline{\bf IV. EWP IN TERMS OF TREE AMPLITUDES}
\bigskip

\begin{table}
\caption{Graphical EWP and tree amplitudes in $\Delta S=1$ $B\to VP$ decays.
Amplitudes for $B^+\to K\pi$ are given for comparison.}
\begin{center}
\begin{tabular}{c c c} \hline \hline
Decay mode  & Tree amplitude & EWP amplitude \\
\hline
$B^+\to\rho^+ K^0$ & $A_V + P_{u,V}$ & $\frac{\kappa}{2}(C_V - 2E_V + 
P_{u,V})$\\
$B^+\to \rho^0 K^+$ & $-\frac{1}{\s}(T_V+C_P+A_V+P_{u,V})$ & 
$\frac{\kappa}{2\s}(3T_P+2C_V+2E_V-P_{u,V})$ \\
$B^0\to \rho^- K^+$ & $-(T_V+P_{u,V})$ & $\frac{\kappa}{2}(2C_V-E_V-P_{u,V})$ \\
$B^0\to\rho^0 K^0$  & $\frac{1}{\s}(-C_P+P_{u,V})$  & $\frac{\kappa}{2\s}
(3T_P+C_V+E_V+P_{u,V})$ \\
 & & \\
$B^+\to K^{*0}\pi^+$ & $A_P+P_{u,P}$ & $\frac{\kappa}{2}(C_P-2E_P+P_{u,P})$ \\
$B^+\to K^{*+}\pi^0$ & $-\frac{1}{\s}(T_P+C_V+A_P+P_{u,P})$ & 
$\frac{\kappa}{2\s}(3T_V+2C_P+2E_P-P_{u,P})$ \\
$B^0\to K^{*+}\pi^-$ & $-(T_P+P_{u,P})$ & $\frac{\kappa}{2}(2C_P-E_P-P_{u,P})$ \\
$B^0\to K^{*0}\pi^0$ & $\frac{1}{\s}(-C_V+P_{u,P})$ & 
$\frac{\kappa}{2\s}(3T_V+C_P+E_P+P_{u,P})$ \\
\hline
$B^+\to K^0\pi^+$ & $A+P_u$ & $\frac{\kappa}{2}(C - 2E + P_u)$\\
$B^+\to K^+\pi^0$ & $-\frac{1}{\s}(T+C+A+P_u)$ & $\frac{\kappa}{2\s}(3T+2C+2E-
P_u)$\\
\hline
\hline
\end{tabular}
\end{center}
\end{table}

\begin{table}
\caption{Graphical tree amplitudes in $\Delta S=0$ $B\to VP$ decays.
Tree amplitude for $B^+\to\pi^+\pi^0$ is given for comparison.}
\begin{center}
\begin{tabular}{c c} \hline \hline
Decay mode  & Tree amplitude \\
\hline
$B^+\to\rho^+ \pi^0$ & $-\frac{1}{\s}(T_P+C_V+P_{u,P}-P_{u,V}+A_P-A_V)$ \\
$B^+\to \rho^0 \pi^+$ & $-\frac{1}{\s}(T_V+C_P-P_{u,P}+P_{u,V}-A_P+A_V)$ \\
$B^+\to K^{*+}\bar K^0$ & $A_V+P_{u,V}$ \\
$B^+\to\bar K^{*0} K^+$ & $A_P+P_{u,P}$ \\
 & \\
$B^0\to \rho^-\pi^+$ & $-(T_V+P_{u,V}+E_P+\frac{1}{2}PA_{u,P}+\frac{1}{2}
PA_{u,V})$ \\
$B^0\to \rho^+\pi^-$ & $-(T_P+P_{u,P}+E_V+\frac{1}{2}PA_{u,P}+
\frac{1}{2} PA_{u,V})$ \\
$B^0\to \rho^0\pi^0$ & $\frac{1}{2}(P_{u,P}+P_{u,V}-C_P-C_V+E_P+E_V+PA_{u,P}+
PA_{u,V})$ \\
$B^0\to K^{*+}K^-$ & $-(E_V+\frac{1}{2} PA_{u,P}+\frac{1}{2} PA_{u,V})$ \\
$B^0\to K^{*-}K^+$ & $-(E_P+\frac{1}{2} PA_{u,P}+\frac{1}{2} PA_{u,V})$ \\
$B^0\to K^{*0}\bar K^0$ & $P_{u,V}+\frac{1}{2} PA_{u,P}+\frac{1}{2} PA_{u,V}$ \\
$B^0\to\bar K^{*0}K^0$ & $P_{u,P}+\frac{1}{2} PA_{u,P}+\frac{1}{2} PA_{u,V}$ \\
\hline
$B^+\to\pi^+\pi^0$ & $-\frac{1}{\s}(T+C)$ \\
\hline
\hline
\end{tabular}
\end{center}
\end{table}

The expressions (\ref{Sreduced})--(\ref{Areduced}), for tree amplitudes 
corresponding to given SU(3) representations, and the 
proportionality relations (\ref{15})(\ref{6}) and (\ref{3}),
can be used with Table I in order to calculate EWP contributions to 
$B\to VP$ decays in terms of graphical tree amplitudes. The results 
for $\Delta S=1$ processes, multiplying $\lambda^{(s)}_t$, are summarized in 
Table II. 
Also included are expressions for the corresponding tree amplitudes 
multiplying $\lambda^{(s)}_u$. For comparison with $B\to PP$ decays, we give 
expressions for the
amplitudes of $B^+\to K^0\pi^+$ and $B^+\to K^+\pi^0$. In Table III we list 
the graphical expansion of tree amplitudes in $\Delta S=0$ $B\to VP$ decays.
For comparison with $B\to PP$, we also include the tree amplitude of 
$B^+\to\pi^+\pi^0$. EWP contributions in this process \cite{GPY}, as well as in
several $VP$ amplitudes involving $T_P$ and $T_V$, are negligible.

Before discussing a few interesting relations between EWP and tree amplitudes
following from Tables II and III, let us recall the relation between our present
results and the traditional approach to EWP contributions.
The graphical EWP amplitudes, which in the conventional approach are 
independent parameters, are given here in terms of graphical tree amplitudes.
In the notation of \cite{GHLRP}, expanded in the case of rescattering effects
to a set of eight graphical EWP amplitudes \cite{EWE}, one finds for $M=P, V$
\bea\label{EW1}
EW_M &=& -\frac{3\kappa}{2}(T_M + P_{u,M'})~,~~~M'\ne M~,\\
\label{EW2}
EW^C_M &=& \frac{3\kappa}{2}(P_{u,M} - C_M)~,\\
\label{EW3}
EWE_M &=& \frac{3\kappa}{2}(P_{u,M} - E_M)~,\\
\label{EW4}
EWA_M &=& \frac{3\kappa}{2}(PA_{u,M} - A_M)~.
\eea
The amplitudes $EWA_M$ do not occur in the processes of Table II. They do 
accur in $B_s$ decays.
 
Tables II and III imply a few SU(3) relations between EWP and tree amplitudes
of corresponding $B\to VP$ decay processes, which are similar to the relation
noted recently to hold in $B^+\to K\pi$ decays \cite{NRPL,GPY}. Starting with 
the latter case, and denoting EWP and tree contributions by $EW$ and $TR$,
respectively, we have 
\bea\label{kpi}
EW(K^0\pi^+) + \s EW(K^+\pi^0) &=& -\frac{3\kappa}{2}[TR(K^0\pi^+) + 
\s TR(K^+\pi^0)]\nonumber\\
&=& -\frac{3\kappa}{\s\lambda^{(d)}_u}A(\pi^+\pi^0) = \frac{3\kappa}{2}(T+C)~.
\eea
This relation follows directly from Eq.~(\ref{15}). The two states, $|K^0\pi^+
\rangle + \s|K^+\pi^0\rangle = |I=3/2\rangle$ and $\s|\pi^+\pi^0\rangle_{\rm 
S-wave}= |I=2\rangle$, are members of a ${\bf 27}$ representation to which
only the $\overline{\bf 15}$ operator contributes. The corresponding relation
in $B\to VP$ is 
\bea\label{VPch}
EW(\rho^+K^0) + \s EW(\rho^0 K^+) + EW(K^{*0}\pi^+) + \s EW(K^{*+}\pi^0) 
\nonumber\\
= -\frac{3\kappa}{\s\lambda^{(d)}_u}[A(\rho^+\pi^0) + A(\rho^0\pi^+)] = 
\frac{3\kappa}{2}(T_P+T_V+C_P+C_P)~.
\eea
In this case the two SU(3)-symmetrized $VP$ states, $|\rho^+K^0\rangle +$
$\s |\rho^0 K^+\rangle + |K^{*0}\pi^+\rangle + \s|K^{*+}\pi^0\rangle$ (isospin 
$3/2$) and $\s(|\rho^+ \pi^0\rangle + |\rho^0\pi^+\rangle)$ (isospin $2$), 
belong to a ${\bf 27}$ representation.

Two other relations can be obtained from Table II
\bea\label{rhoK}
EW(\rho^+K^0) + \s EW(\rho^0 K^+) &=& -\frac{3\kappa}{2}[TR(K^{*0}\pi^+) + 
\s TR(K^{*+}\pi^0)]\nonumber\\
&=& \frac{3\kappa}{2}(T_P+C_V)~,\\
\label{K*pi}
EW(K^{*0}\pi^+) + \s EW(K^{*+}\pi^0) &=& -\frac{3\kappa}{2}[TR(\rho^+ K^0) + 
\s TR(\rho^0 K^+)]\nonumber\\
&=& \frac{3\kappa}{2}(T_V+C_P)~.
\eea
These relations can be understood in the following way. The two $I=3/2$ states, 
$|\rho^+K^0\rangle + \s |\rho^0 K^+\rangle$ and $|K^{*0}\pi^+\rangle + 
\s|K^{*+}\pi^0\rangle$, form the sum and difference, respectively, of a ${\bf 
27}$ and a ${\bf 10}$ representation. This can be easily verified in Table I.
The ${\bf 27}$ and ${\bf 10}$ states obtain contributions only from 
$\overline{\bf 15}$ and ${\bf 6}$ operators, respectively.
Eqs.~(\ref{15}) and (\ref{6}), in which the proportionality constants have equal 
magnitudes and opposite signs, lead immediately to Eqs.~(\ref{rhoK}) and 
(\ref{K*pi}). It is clear from these considerations that these relations hold 
also in the presence of current-current and EWP $\bar b\to\bar q c\bar c$ 
operators transforming as a antitriplet.

A useful approximation is obtained by neglecting in $B^+ \to\rho\pi$ the 
rescattering amplitudes, $P_{u,M}+A_M$ \cite{resc}. The smallness
of these terms can be tested in $B^+\to K^{*+}\bar K^0$ and $B^+\to\bar K^{*0} 
K^+$. In this approximation, one can also express the sums of EWP contributions 
in (\ref{rhoK}) and (\ref{K*pi}) in terms of $B^+\to\rho\pi$ amplitudes, 
similar to (\ref{kpi})
\bea
EW(\rho^+K^0) + \s EW(\rho^0 K^+) &=&  -\frac{3\kappa}{\s\lambda^{(d)}_u}
A(\rho^+\pi^0)~,\\
EW(K^{*0}\pi^+) + \s EW(K^{*+}\pi^0) &=& -\frac{3\kappa}{\s\lambda^{(d)}_u}
A(\rho^0\pi^+)~.
\eea
These approximate relations are useful when studying charge-averaged ratios 
of rates for the processes on the left hand sides.

Relations of the form (\ref{rhoK}) and (\ref{K*pi}) are obeyed also by EWP and 
tree decay amplitudes of $B^0$ to $\rho^-K^+,~\rho^0K^0,~K^{*+}\pi^-$ and 
$K^{*0}\pi^0$. This is easy to understand. These contributions, as well as the 
entire decay amplitudes, which also contain dominant gluonic penguin terms, 
satisfy an 
isospin relation with corresponding $B^+$ decay amplitudes \cite{iso}
\beq
A(B^+\to\rho^+K^0) + \s A(B^+\to\rho^0K^+) = A(B^0\to\rho^-K^+) + 
\s A(B^0\to\rho^0K^0)~.
\eeq
An analogous isospin equality holds for $B\to K^*\pi$ decay amplitudes. Finally,
a relation similar to Eq.~(\ref{VPch}), between $\Delta S=1$ decays to $I=3/2$ 
on the one hand
and $\Delta S=0$ decays to $I=2$ on the other hand,  can be written by combining all seven 
neutral $B$ decay amplitudes to $\rho K,~K^*\pi$ and $\rho\pi$ states.
\bea\label{VPneut}
EW(\rho^-K^+) + \s EW(\rho^0 K^0) + EW(K^{*+}\pi^-) + \s EW(K^{*0}\pi^0) 
\nonumber\\
= -\frac{3\kappa}{2\lambda^{(d)}_u}[A(\rho^-\pi^+) + A(\rho^+\pi^-) + 
2A(\rho^0\pi^0)] = \frac{3\kappa}{2}(T_P+T_V+C_P+C_P)~.
\eea

\bigskip
\centerline{\bf V. APPLICATIONS}
\bigskip

{\it 1. Resolving EWP in $B^+\to K\pi$}
\medskip

Let us first reiterate the manner in which Eq.~(\ref{kpi}) has been applied in 
order to obtain a model-independent constraint on $\gamma$ from the 
charge-averaged ratio \cite{NRPL}
\beq\label{R*}
R^{-1}_*(K\pi)\equiv \frac{2[B(B^+\to K^+\pi^0) + B( B^-\to K^-\pi^0)]}
{B(B^+\to K^0\pi^+) + B(B^-\to \bar K^0\pi^-)}~.
\eeq
Using our graphical notation for amplitudes, one has
\bea
\s A(B^+\to K^+\pi^0) &=& -\lambda_u^{(s)}[T+C+P_{uc}+A] -
\lambda_t^{(s)}[P_{tc}-\s EW(K^+\pi^0)]~,\nonumber\\
A(B^+\to K^0\pi^+) &=& \lambda_u^{(s)}[P_{uc}+A] +
\lambda_t^{(s)}[P_{tc}+EW(K^0\pi^+)]~.
\eea
The two electroweak penguin terms, containing also contributions from $\bar 
b\to\bar q c\bar c$ operators, satisfy Eq.~(\ref{kpi}).
Substituting these expressions into (\ref{R*}), applying unitarity of the CKM 
matrix, and expanding in small quantities, one finds
\beq\label{R*th}
R^{-1}_*(K\pi) = 1 - 2\epsilon \cos\phi (\cos\gamma - \delta_{EW}) + 
{\cal O}(\epsilon^2) + {\cal O}(\epsilon\epsilon_A) + {\cal O}(\epsilon^2_A)~,
\eeq
where $\phi={\rm Arg}\left ([T+C]/[P_{tc}+EW(B^+\to K^0\pi^+)]\right )$. 

The real and positive parameter $\delta_{EW}$ \cite{NRPL} stands for the ratio 
of EWP and tree 
contributions in the sum $A(B^+\to K^0\pi^+)+\s A(B^+\to K^+\pi^0)$, and is 
determined purely by Wilson coefficients and by a presently poorly known CKM 
factor [see Eq.~(\ref{kpi})]
\beq\label{deltaEW}
\delta_{EW} = -\frac{3\kappa}{2}|\frac{V^*_{cb}V_{cs}}{V^*_{ub}V_{us}}| 
=0.65\pm 0.15~.
\eeq
The quantity
$\epsilon=[|V^*_{ub}V_{us}|/|V^*_{cb}V_{cs}|][|T+C|/|P_{tc}+EW(B^+\to 
K^0\pi^+)|]$ is measurable from \cite{GRL,CLEO2}
\bea
\epsilon = \sqrt2 \frac{V_{us}}{V_{ud}}\frac{f_K}{f_\pi}
\frac{|A(B^+\to \pi^0\pi^+)|}{|A(B^+\to K^0\pi^+)|}= 0.21\pm 0.05~.
\eea
$\epsilon_A$ denotes a small rescattering amplitude \cite{resc}, which 
introduces a term $P_{uc}+A$ with weak phase $\gamma$ into the 
$B^+\to K^0\pi^+$ decay amplitude. Keeping the dominant term in (\ref{R*th}) 
and neglecting smaller terms, one obtains the bound
\beq\label{kpibound}
|\cos\gamma - \delta_{EW}| \ge \frac{|1-R^{-1}_*(K\pi)|}{2\epsilon}~.
\eeq
This bound can provide useful information about $\gamma$ in case that a 
value different from one is measured for $R^{-1}_*$. Further information about
the weak phase can be obtained by measuring separately $B^+$ and $B^-$ decay 
rates \cite{NRPRL}. Eq.~(\ref{kpi}) plays a crucial role in these applications.

\bigskip
{\it 2. Generalization to $B^+\to \rho K$ and $B^+\to K^*\pi$}
\medskip

Whereas Eq.~(\ref{kpi}) relates EWP and tree contributions in the same sum of 
two $B^+\to K\pi$ amplitudes, the analogous Eqs.~(\ref{rhoK}) and (\ref{K*pi}) 
relate EWP contributions in a sum of $B^+\to\rho K$ amplitudes to tree 
contributions in another sum of $B\to K^*\pi$ amplitudes, and vice versa.
This introduces some hadronic dependence in possible constraints on
$\gamma$ from these processes. 

Consider, for instance, the charge-averaged ratio of rates for the
processes $B^{\pm}\to\rho^{\pm} K^0$ and $B^{\pm}\to\rho^0 K^{\pm}$ 
\beq\label{R*V}
R^{-1}_*(\rho K)\equiv \frac{2[B(B^+\to \rho^0 K^+) + B( B^-\to \rho^0 K^-)]}
{B(B^+\to \rho^+K^0) + B(B^-\to \rho^-\bar K^0)}~.
\eeq
Using Table II for graphical expressions of amplitudes, applying 
Eq.~(\ref{rhoK}), and neglecting rescattering contributions $P_{uc,V}+A_V$,
which affect the above ratio only by second order terms, as in (\ref{R*th}), 
one has
\bea
\s A(B^+\to \rho^0 K^+) &=& -|\lambda_u^{(s)}|[(T_V+C_P)
e^{i\gamma}-(T_P+C_V)\delta_{EW}] -
\lambda_t^{(s)}[P_{tc,V}+ EW]~,\nonumber\\
A(B^+\to \rho^+K^0) &=& \lambda_t^{(s)}[P_{tc,V}+EW]~,~~~~~~~~EW\equiv
EW(B^+\to \rho^+K^0)~.
\eea
We define two ratios of amplitudes
\beq
\epsilon_Ve^{i\phi_V}=\frac{|V^*_{ub}V_{us}|}{|V^*_{cb}V_{cs}|}
\frac{T_V+C_P}{P_{tc,V}+EW}~,
~~~~~~~~~~\epsilon_Pe^{i\phi_P}=\frac{|V^*_{ub}V_{us}|}{|V^*_{cb}V_{cs}|}
\frac{T_P+C_V}{P_{tc,V}+EW}~,
\eeq
the magnitudes of which are measured in $B^+\to\rho^0\pi^+$ and $B^+\to\rho^+
\pi^0$, respectively (see Table III, where we neglect rescattering terms 
$P_{uc,M}+A_M$)
\beq
\epsilon_V = \sqrt2 \frac{V_{us}}{V_{ud}}\frac{f_K}{f_\pi}
\frac{|A(B^+\to \rho^0\pi^+)|}{|A(B^+\to \rho^+K^0)|}~,~~~~~~~~~~
\epsilon_P = \sqrt2 \frac{V_{us}}{V_{ud}}\frac{f_{K^*}}{f_\rho}
\frac{|A(B^+\to \rho^+\pi^0)|}{|A(B^+\to \rho^+K^0)|}~.
\eeq
We find
\beq
R^{-1}_*(\rho K) = 1 - 2\epsilon_V\cos\phi_V\cos\gamma +
2\epsilon_P\delta_{EW}\cos\phi_P~.
\eeq

This expression, which neglects higher order corrections,  simplifies into the 
form (\ref{R*th}) in the case $T_V+C_P=T_P+C_V$, or
$A(B^+\to\rho^0\pi^+)=A(B^+\to\rho^+\pi^0)$. In general, this is not the case.
Without making any assumption about the magnitudes of these amplitudes and 
their strong phases, one obtains the rather weak constraint
\beq
|\cos\gamma| \ge \frac{|1-R^{-1}_*(\rho K)|}{2\epsilon_V} - \delta_{EW}\left(
\frac{\epsilon_P}{\epsilon_V}\right )~.
\eeq
This bound is manifestly weaker than the constraint (\ref{kpibound}) obtained
\cite{NRPL} for $B\to K\pi$. In order to exclude values of $\gamma$ around 
$90^\circ$ the right hand side must be positive.
This does not only require that a value different from one is measured for 
$R^{-1}_*(\rho K)$, but also that $|A(B^+\to \rho^+\pi^0)|$ is considerably 
smaller than $|A(B^+\to \rho^0\pi^+)|$. A similar argument applies to 
the ratio $R^{-1}_*(K^{*}\pi)$ in $B^{\pm}\to K^{*}\pi$ decays. Since these
two ratios have not yet been measured, no constraint on $\gamma$ can be 
obtained at this time.

\bigskip 
{\it 3. A lower bound on $\gamma$ from $B^0\to K^{*+}\pi^-$ 
and $B^+\to \phi K^+$}
\medskip

A plausible argument which favors $\cos\gamma <0$ was presented recently in 
ref.~\cite{VP2}, based primarily on recent CLEO data on $B^0\to K^{*+}\pi^-$ 
and $B^+\to \phi K^+$. In this argument, only the dominant amplitudes 
contributing to these processes, $P_{tc,P},~EW_P$ and $T_P$ , were taken into 
account, 
while smaller terms were neglected. The argument was based on certain model 
calculations of EWP amplitudes \cite{RFDH,ALI}, which imply $EW_P \approx 
P_P/2$. This relation, obtained for certain values of a set of parameters, 
including the effective number of colors in a $1/N_c$ expansion, also assumes 
that the two amplitudes have equal strong phases.
Here we would like to replace these model-dependent assumptions by our
general SU(3) results, which relate electroweak penguin contributions to tree
amplitudes rather than to gluonic penguin amplitudes. As in \cite{VP2}, 
we will keep only the dominant and subdominant terms.

The amplitude of $B^0\to K^{*+}\pi^-$ is
\beq\label{k*pi}
A(B^0\to K^{*+}\pi^-) = -\lambda^{(s)}_u[T_P + P_{uc,P}] -\lambda^{(s)}_t
[P_{tc,P} - EW(K^{*+}\pi^-) - EW_c(K^{*+}\pi^-)]~,
\eeq
where $EW(K^{*+}\pi^-)$ is given in Table II
\beq\label{EWk*}
EW(K^{*+}\pi^-) = \frac{\kappa}{2}(2C_P - E_P - P_{u,P})~.
\eeq
The amplitude of $B^+\to\phi K^+$ involves also the SU(3) singlet component of
the $\phi$. In a general SU(3) analysis, this component introduces three new 
reduced SU(3) amplitudes, of the $\overline{\bf 3},~{\bf 6}$ and 
$\overline{\bf 15}$ 
operators, for the final octet state. These three amplitudes are described by 
three new graphs: A disconnected penguin diagram, $S_P$ \cite{VP2}, in which 
a singlet $q\bar q$ pair is connected to the rest of the diagram by at least
three gluons, and two ``hairpin" diagrams, of annihilation ($AS_P$) and 
exchange ($ES_P$) types, in which the extra $q\bar q$ forms the singlet 
vector meson. Thus, one has
\beq\label{kphi}
A(B^+\to\phi K^+) = \lambda^{(s)}_u[A_P+P_{uc,P}+AS_P]
+ \lambda^{(s)}_t[P_{tc,P}+S_P+EW(\phi K^+)+EW_c(\phi K^+)]~,
\eeq
where, applying (\ref{EW1})--(\ref{EW3}), 
\beq\label{EWphi}
EW(\phi K^+) = -\frac{1}{3}(EW_P+EW^C_P-2EWE_P)=\frac{\kappa}{2}(T_P+C_P-
2E_P+P_{u,P}+P_{u,V})~.
\eeq

We will assume, as usual \cite{BBNS,GHLR,GHLRP}, that $T_P$ is larger than all 
other tree amplitudes 
and larger than the current-current amplitude associated with 
$\bar b\to \bar q c\bar c$. (Recall that the CKM coefficients are factored 
out).
Similarly, we will assume that $|EW| \gg |EW_c|$. The amplitude $S_P$ will 
be neglected by virtue of the Okubo-Zweig-Iizuka (OZI) rule. We note that
in the factorization approach \cite{ALI}, $S_P$ is very sensitive to the number
of colors $N_c$, and vanishes at $N_c=3$. 
The EWP contribution in $B^+\to\phi K^+$ is dominated by a term 
proportional to $T_P$, which is measured in
$B^+\to\rho^+\pi^0$ and $B^0\to\rho^+\pi^-$ as discussed below.

Keeping only dominant and subdominant terms in each amplitude, one has
\bea\label{K*}
A(B^0\to K^{*+}\pi^-) &=& -\lambda^{(s)}_t\,P_{tc,P} - \lambda^{(s)}_u\,T_P~,\\ 
\label{phi}
A(B^+\to \phi K^+) &=& \lambda^{(s)}_t\,[P_{tc,P}+\frac{\kappa}{2}T_P]~.
\eea
In this approximation, the two amplitudes, $P_{tc,P}$ and $T_P$, contribute 
with the same weak phase in $B^+\to \phi K^+$, and interfere
with a relative weak phase $\pi-\gamma$ in $B^0\to K^{*+}\pi^-$.
Defining 
\beq
r\,e^{i\delta} = \frac{|\lambda^{(s)}_u|}{|\lambda^{(s)}_t|}\frac{T_P}
{P_{tc,P}}~,~~~~~(r>0)~,
\eeq
we have
\bea\label{a}
A(B^0\to K^{*+}\pi^-) &=& -\lambda^{(s)}_t\,P_{tc,P}\,[1-r\,e^{i(\delta+
\gamma)}]~, \\ 
\label{b}
A(B^+\to \phi K^+) &=& \lambda^{(s)}_t\,P_{tc,P}\,[1-
\frac{1}{3}\delta_{EW}\,r\,e^{i\delta}]~,
\eea
where $\delta_{EW}$ is defined in (\ref{deltaEW}).

In the limit of neglecting the tree amplitude, $r=0$, the rates of the two 
processes are seen to be equal. Experiments obtain 90$\%$ confidence level 
limits on the charge-averaged rates \cite{CLEO}, $\b(B^0\to K^{*\pm}\pi^{\mp})>
12\times 10^{-6}$ and $\b(B^{\pm}\to \phi K^{\pm})<5.9\times 10^{-6}$.
This is evidence for a nonzero contribution of $T_P$, namely $r\ne 0$. The ratio
of charge-averaged rates satisfies, at 90$\%$ c.l. (we neglect the $B^+-B^0$ 
lifetime difference)
\beq\label{ratio}
\frac{|A(B^0\to K^{*\pm}\pi^{\mp})|^2}{|A(B^{\pm}\to \phi K^{\pm})|^2} =
\frac{1+r^2-2r\,\cos\delta\cos\gamma}{1+(\delta_{EW}/3)^2\,r^2-(2/3)\,
\delta_{EW}\,r\,\cos\delta} > 2.0~.
\eeq

In order to use this inequality for information about $\gamma$, one must 
include some input about $r$ and $\delta$, the relative magnitude and 
strong phase 
of tree and penguin amplitudes in $B^0\to K^{*+}\pi^-$. A reasonable assumption, 
supported both by perturbative \cite{BBNS,BSS} and statistical \cite{SW} 
calculations, is that $\delta$ does not exeed $90^\circ$, i.e. $\cos\delta 
\ge 0$. A conservative assumption about $r$ is $r\le 1$. Making these two
assumptions, one finds 
\beq\label{gamma}
\cos\gamma -\frac{2}{3}\delta_{EW} < \frac{-1 + r^2[1-2(\delta_{EW}/3)^2]}{2r}~.
\eeq
This implies $\gamma > 62^\circ$ for $r=1$, and $\gamma > 105^\circ$ for 
$r=0.5$, when $\delta_{EW}$ is taken in the range (\ref{deltaEW}).
Some very indirect evidence for $r<0.55$ was presented in \cite{VP2}, 
relying on a nonzero value of $T_P$ obtained from $B^0\to\rho^{\pm}\pi^{\mp}$ 
and $B^+\to\rho^0\pi^+/\omega\pi^+$. More direct information 
about $r$ is required, and can be inferred from future rate measurements of 
$B^+\to\rho^+\pi^0$ or $B^0\to\rho^+\pi^-$ and $B^+\to K^{*0}\pi^+$.  These 
processes are dominated by $T_P$ 
and $P_{tc,P}$, respectively (see Table III and \cite{VP2}). 

The bound on $\gamma$ (\ref{gamma}), which is based on the experimental limit 
(\ref{ratio}), neglects smaller terms in the amplitudes (\ref{k*pi}) and
(\ref{kphi}), primarily the color-suppressed terms $C_P$ in 
(\ref{EWk*}) and (\ref{EWphi}) and the OZI-suppressed penguin amplitude $S_P$
in (\ref{kphi}). For $|C_P/T_P|=0.1~(0.2)$ \cite{BBNS},  
our limits move up or down by about $5^\circ~(10^\circ)$, depending on whether 
the interference between $C_P$ and $T_P$ is destructive or constructive, 
respectively. 

The above limits also assume [by SU(3)] equal gluonic penguin contributions in 
the two processes. An important question relevant to these bounds is the 
magnitude and sign of SU(3) breaking in penguin amplitudes. For instance,
if the penguin amplitude in $B^+\to \phi K^+$ is {\it smaller} by $30\%$ than 
in $B^0\to K^{*+}\pi^-$, then the above bounds are completely invalidated. 
On the other hand, the constraint becomes stronger if the penguin amplitude in 
$B^+\to \phi K^+$ is {\it larger} than in $B^0\to K^{*+}\pi^-$. This is the 
case in explicitly SU(3) breaking factorization-based calculations, 
in which the two amplitudes involve the products of corresponding vector 
meson decay constants and $B$-to-pseudoscalar form factors.

In the factorization approximation, SU(3) breaking factors in penguin and tree 
amplitudes occuring in  Eqs.~(\ref{K*}) and (\ref{phi}) are given by \cite{ALI}
\beq
R_{SU(3)}=
\frac{P_{tc,P}(B^+\to \phi K^+)}{P_{tc,P}(B^0\to K^{*+}\pi^-)} =
\frac{T_P(B^+\to \phi K^+)}{T_P(B^0\to K^{*+}\pi^-)}\simeq
\frac{f_{\phi}}{f_{K^*}}\frac{F_{BK}(m^2_\phi)}{F_{B\pi}(m^2_{K^*})}
\simeq 1.25~.
\eeq
Since this factor enhances the amplitude of $B^+\to \phi K^+$ 
relative to that of $B^0\to K^{+}\pi^-$, the bound on $\gamma$ (\ref{ratio}) 
becomes stronger by a factor $R^2_{SU(3)}$. This would imply, for instance, 
$\gamma > 80^{\circ}$ if a value $r=1$ is measured in $B^+\to\rho^+\pi^0$ and 
$B^+\to K^{*0}\pi^+$, and a stronger bound if a smaller value of $r$ is
measured. This, and the above comment on the possibility that
$R_{SU(3)}<1$, illustrate the sensitivity of these bounds to SU(3) breaking
effects.

Although the present experimental inequality (\ref{ratio}) (which may change 
with time) is already interesting,
our above discussion shows that it would be premature at this point to 
translate this inequality into a realistic lower bound on $\gamma$. Further 
study is required of the following effects: 
\begin{itemize}
\item SU(3) breaking in penguin amplitudes: are these amplitudes 
approximately factorizable? 
\item Magnitudes and strong phases of smaller terms, including 
color-suppressed tree and OZI-suppressed penguin amplitudes.
\item An actual measurement of $r$, the ratio of tree-to-penguin amplitudes
in $B^0\to K^{*+}\pi^-$.
\end{itemize}    

\newpage
\centerline{\bf VI. CONCLUSION}
\bigskip

We have studied EWP amplitudes in $B\to VP$ decays within the model independent
framework of flavor SU(3). While retaining only contributions from the
dominant $(V-A)(V-A)$ operators, $Q_9$ and $Q_{10}$, we were able to express
these contributions in terms of tree amplitudes. This reduces considerably 
the number of hadronic parameters describing a large number of processes. 

Two applications were demontrated in attempting to constrain the weak phase 
$\gamma$. In $B^+\to \rho^+K^0$ and $B^+\to\rho^0 K^+$ (or in $B^+\to K^{*0}
\pi^+$ and $B^+\to K^{*+}\pi^0$) we studied a generalization of the method 
suggested in \cite{NRPL} for $B^+\to K\pi$. We find that the constraint 
becomes weaker due to some dependence on hadronic matrix elements.

In a second application we reexamined the decays $B^0\to K^{*+}\pi^-$ and 
$B^+\to \phi K^+$, studied recently in \cite{VP2}, where EWP contributions
were taken from model-calculations. We kept only the 
dominant and subdominant terms and assumed that the relevant strong phase 
does not exceed
$90^\circ$. The present lower limit on the charge-averaged ratio of rates for 
these two processes leads to an interesting lower bound on $\gamma$, 
Eq.~(\ref{gamma}). The bound depends on $r$, the ratio of tree to penguin 
amplitudes in $B^0\to K^{*+}\pi^-$, which can be measured in 
$B^{+,0}\to\rho^+\pi^{0,-}$ and $B^+\to K^{*0}\pi^+$. Corrections from 
color-suppressed and OZI-suppressed terms are estimated to move the bound by 
about 10 degrees. A 
larger correction may be due to SU(3) breaking in penguin 
amplitudes. In case that SU(3) breaking decreases the penguin 
amplitude in $B^+\to \phi K^+$ relative to the one in $B^0\to K^{*+}\pi^-$,
contrary to the prediction of factorization, the bound on $\gamma$ may become 
considerably weaker. A proof of approximate factorization for penguin 
amplitudes in $B\to VP$ decays, which would strengthen the bound, is 
therefore of great importance. 

[{\it Note added}: Three months after the submission for publication of this 
paper a work appeared \cite{DXJ}, in which flavor SU(3) symmetry (or, actually 
an extended nonet symmetry) was applied to charmless $B\to VP$ decays, in order 
to prove several relations between CP violating rate differences. This work did 
not make use of the symmetry relations between EWP and tree amplitudes 
studied in the present paper.]  
 
\bigskip
\centerline{\bf ACKNOWLEDGMENTS}
\bigskip

I thank Dan Pirjol and Jon Rosner for useful discussions.  
This work was supported in part by the Israel Science Foundation founded by the 
Israel Academy of Sciences and Humanities, by the United States -- Israel 
Binational Science Foundation under Research Grant Agreement 98-00237, and by 
the fund for the promotion of research at the Technion. 
\bigskip

\newpage

\def \ajp#1#2#3{Am. J. Phys. {\bf#1}, #2 (#3)}
\def \apny#1#2#3{Ann. Phys. (N.Y.) {\bf#1}, #2 (#3)}
\def \app#1#2#3{Acta Phys. Polonica {\bf#1}, #2 (#3)}
\def \arnps#1#2#3{Ann. Rev. Nucl. Part. Sci. {\bf#1}, #2 (#3)}
\def \art{and references therein}
\def \cmts#1#2#3{Comments on Nucl. Part. Phys. {\bf#1}, #2 (#3)}
\def \cn{Collaboration}
\def \cp89{{\it CP Violation,} edited by C. Jarlskog (World Scientific,
Singapore, 1989)}
\def \epjc#1#2#3{Euro.~Phys.~J.~C {\bf #1}, #2 (#3)}
\def \epl#1#2#3{Europhys.~Lett.~{\bf #1}, #2 (#3)}
\def \ib{{\it ibid.}~}
\def \ibj#1#2#3{~{\bf#1}, #2 (#3)}
\def \ijmpa#1#2#3{Int. J. Mod. Phys. A {\bf#1}, #2 (#3)}
\def \jpb#1#2#3{J.~Phys.~B~{\bf#1}, #2 (#3)}
\def \jhep#1#2#3{JHEP {\bf#1}, #2 (#3)}
\def \mpla#1#2#3{Mod. Phys. Lett. A {\bf#1}, #2 (#3)}
\def \nc#1#2#3{Nuovo Cim. {\bf#1}, #2 (#3)}
\def \np#1#2#3{Nucl. Phys. {\bf#1}, #2 (#3)}
\def \pisma#1#2#3#4{Pis'ma Zh. Eksp. Teor. Fiz. {\bf#1}, #2 (#3) [JETP Lett.
{\bf#1}, #4 (#3)]}
\def \pl#1#2#3{Phys. Lett. {\bf#1}, #2 (#3)}
\def \pla#1#2#3{Phys. Lett. A {\bf#1}, #2 (#3)}
\def \plb#1#2#3{Phys. Lett. B {\bf#1}, #2 (#3)}
\def \pr#1#2#3{Phys. Rev. {\bf#1}, #2 (#3)}
\def \prc#1#2#3{Phys. Rev. C {\bf#1}, #2 (#3)}
\def \prd#1#2#3{Phys. Rev. D {\bf#1}, #2 (#3)}
\def \prl#1#2#3{Phys. Rev. Lett. {\bf#1}, #2 (#3)}
\def \prp#1#2#3{Phys. Rep. {\bf#1}, #2 (#3)}
\def \ptp#1#2#3{Prog. Theor. Phys. {\bf#1}, #2 (#3)}
\def \ptwaw{Plenary talk, XXVIII International Conference on High Energy
Physics, Warsaw, July 25--31, 1996}
\def \rmp#1#2#3{Rev. Mod. Phys. {\bf#1}, #2 (#3)}
\def \rp#1{~~~~~\ldots\ldots{\rm rp~}{#1}~~~~~}
\def \stone{{\it $B$ Decays} (Revised 2nd Edition), edited by S. Stone
(World Scientific, Singapore, 1994)}
\def \yaf#1#2#3#4{Yad. Fiz. {\bf#1}, #2 (#3) [Sov. J. Nucl. Phys. {\bf #1},
#4 (#3)]}
\def \zhetf#1#2#3#4#5#6{Zh. Eksp. Teor. Fiz. {\bf #1}, #2 (#3) [Sov. Phys. -
JETP {\bf #4}, #5 (#6)]}
\def \zpc#1#2#3{Zeit. Phys. C {\bf#1}, #2 (#3)}
\def \zpd#1#2#3{Zeit. Phys. D {\bf#1}, #2 (#3)}


\begin{thebibliography}{99}
\bibitem{EWE} The other four graphical amplitudes will be denoted by $EWE_M$ 
and $EWA_M$ ($M=V, P$). In $EWE_M$ the spectator $q'$ participates in the 
transition $\bar b\to \bar q q' \bar q'$ while the other $q'$ enters the meson 
$M$. In $EWA_M$ the spectator $q$ participates in the transition and $q'$ 
(rather than $\bar q'$) enters $M$.
\end{thebibliography}
\end{document}